\documentclass[twocolumn,showpacs,superscriptaddress,amsmath,amssymb,aps,pra]{revtex4-1}
\usepackage{bm}
\usepackage{graphicx}
\usepackage{dcolumn}
\usepackage{amsmath,txfonts}
\usepackage{epstopdf}
\usepackage[mathlines]{lineno}
\usepackage{color}
\usepackage[colorlinks=true,linkcolor=blue,citecolor=blue]{hyperref}
\usepackage[normalem]{ulem}

\begin{document}

\title{Dispersive readout of a weakly coupled qubit via the parity-time-symmetric phase transition}

\author{Guo-Qiang Zhang}
\affiliation{Interdisciplinary Center of Quantum Information and Zhejiang Province Key Laboratory of Quantum Technology and Device, Department of Physics and State Key Laboratory of Modern Optical Instrumentation, Zhejiang University, Hangzhou 310027, China}
\affiliation{Quantum Physics and Quantum Information Division, Beijing Computational Science Research Center, Beijing 100193, China}

\author{Yi-Pu Wang}
\affiliation{Interdisciplinary Center of Quantum Information and Zhejiang Province Key Laboratory of Quantum Technology and Device, Department of Physics and State Key Laboratory of Modern Optical Instrumentation, Zhejiang University, Hangzhou 310027, China}

\author{J. Q. You}
\email{jqyou@zju.edu.cn}
\affiliation{Interdisciplinary Center of Quantum Information and Zhejiang Province Key Laboratory of Quantum Technology and Device, Department of Physics and State Key Laboratory of Modern Optical Instrumentation, Zhejiang University, Hangzhou 310027, China}

\begin{abstract}
For some cavity-quantum-electrodynamics systems, such as a single electron spin coupled to a passive cavity, it is challenging to reach the strong-coupling regime. In such a weak-coupling regime, the conventional dispersive readout technique cannot be used to resolve the quantum states of the spin. Here we propose an improved dispersive readout method to measure the quantum states of a weakly coupled qubit by harnessing either one or two auxiliary cavities linearly coupled to the passive cavity containing the qubit. With appropriate parameters in both cases, the system excluding the qubit can exhibit a parity-time-symmetric phase transition at the exceptional point (EP). Because the EP can amplify the perturbation induced by the qubit and the parity-time symmetry can narrow the linewidths of the peaks in the transmission spectrum of the passive cavity, we can measure the quantum states of the weakly coupled qubit via this transmission spectrum. Owing to the weak coupling between the qubit and the passive cavity, the backaction due to the measurement of the qubit can also be reduced in comparison with the conventional dispersive readout technique in the strong-coupling regime.
\end{abstract}

\date{\today}

\maketitle

\section{Introduction}

The storage, manipulation, and readout of the states of a qubit are basic tasks in quantum information processing~\cite{Nielsen00,Bennett00,Steane98}. In circuit quantum electrodynamics (QED) (see, e.g., \onlinecite{Schoelkopf08,Xiang13,Kurizki15} for reviews), dispersive quantum nondemolition (QND) measurement is implementable for the readout of quantum states of a superconducting (SC) qubit~\cite{Wallraff04}, when it is strongly coupled to a coplanar waveguide resonator, i.e., a one-dimensional (1D) on-chip cavity. With the qubit-cavity system in the dispersive regime, frequency shifts occur for the cavity, depending on the states of the qubit, and can be probed by measuring the transmission spectrum of the cavity.

With technological advancement, the dispersive QND readout has been applied to various solid-state systems, including the ac-driven system~\cite{Kohler17}, spin ensembles~\cite{Haigh15}, and multilevel systems~\cite{Burkard16,Benito16}. Also, it has been extended to the ultrastrong-coupling regime~\cite{Zueco09,Kohler18}. Indeed, strong-~\cite{Wallraff04}, ultrastrong-~\cite{Niemczyk10,Forn10,Forn16,You17}, and even deep-strong-coupling~\cite{Yoshihara16} regimes have been reached in circuit QED systems, but they are difficult to achieve for some other systems such as a single-electron spin coupled to a cavity, which still remains in the weak-coupling regime~\cite{Xiang13,Kurizki15}. Very recently, it was proposed~\cite{Troiani18} to dispersively measure the states of a weakly coupled qubit using a single 2D square SC cavity with a pair of near-resonant modes $A$ and $B$, where both cavity modes are coupled to the probe field but only mode $A$ is weakly coupled to the qubit. In the circuit QED, 1D rather than 2D SC cavities are commonly used because a 1D cavity can yield a smaller effective volume to produce a stronger coupling strength.
In addition, due to the requirement of two near-resonant modes in a 2D cavity, it is difficult to avoid the direct coupling of the probed qubit to the auxiliary mode (i.e., mode $B$)~\cite{Bonizzoni18}. Moreover, the effect of the qubit's decay was neglected in Ref.~\cite{Troiani18}, but in the considered weak-coupling regime, the decay rate of the qubit is comparable to the frequency detuning between the qubit and mode $A$ and thus it cannot be ignored.

A SC circuit with party-time ($\mathcal{PT}$) symmetry was proposed in Ref.~\cite{Quijandria18}, which consists of two cavities with balanced gain and loss. The $\mathcal{PT}$-symmetric system has a non-Hermitian Hamiltonian, but with real energy spectrum in its $\mathcal{PT}$-symmetric phase~\cite{Bender98,Mostafazadeh02-1,Mostafazadeh02-2,Mostafazadeh02-3}. When varying the parameters, the system can experience a phase transition in the parameter space~\cite{Konotop16,Bender05,Bender07}, from the $\mathcal{PT}$-symmetric phase with real eigenvalues to the $\mathcal{PT}$-symmetry-breaking phase with complex eigenvalues. The corresponding critical point is called the $n$th-order exceptional point (EP$_{n}$)~\cite{Heiss12}, where $n$ modes become coalescent. The EPs have been widely studied in various $\mathcal{PT}$-symmetric systems, owing to their intriguing properties, such as the unidirectional invisibility~\cite{Lin11,Feng13}, the lowering of chaos' threshold power~\cite{Lv15}, the induced abnormal laser phenomena~\cite{Liertzer12,Feng14,Hodaei14}, and the enhanced spontaneous emission~\cite{Lin16}.  In particular, the sensitivity of the detection can be enhanced near an EP in, e.g., microcavity sensors~\cite{Wiersig14} and metrology~\cite{Liu16}. Actually, a weak perturbation $\xi$ on the $\mathcal{PT}$-symmetric Hamiltonian can induce a spectral splitting of the non-Hermitian system around an EP$_{n}$ by following a $\xi^{1/n}$ dependence~\cite{Hodaei17,Chen17}. This indicates that a higher-order EP can enhance the sensitivity of the sensors more.

In this paper, we propose two dispersive readout schemes to measure the states of a qubit weakly coupled to a cavity $a$ [cf. Fig.~\ref{fig1}(a)] (i.e., in the weak-coupling regime), by harnessing auxiliary cavities. In the first scheme, an auxiliary cavity $b$ with gain is introduced [see Fig.~\ref{fig1}(b)], which is linearly coupled to cavity $a$. In this scheme, cavity $a$ is far off resonance with the qubit, but on resonance with cavity $b$. With the balanced loss and gain, the $\mathcal{PT}$-symmetric subsystem, consisting of cavities $a$ and $b$, can exhibit a phase transition at an $\rm{EP}_{2}$. We find that the difference of the system's energy spectrum when the qubit is in the ground and excited states, respectively, can be amplified near the $\rm{EP}_{2}$. Also, linewidths of the peaks in the transmission spectrum of cavity $a$ are squeezed. These can be used to probe the states of the qubit in cavity $a$. In the second scheme, we harness two auxiliary cavities $b$ and $c$ [see Fig.~\ref{fig1}(c)]. With appropriate parameters, the subsystem, consisting of the three cavities $a$, $b$, and $c$, can also be $\mathcal{PT}$ symmetric and have an $\rm{EP}_{3}$. Compared with the first scheme, the sensitivity of the detection is enhanced when probing the states of the qubit around the $\rm{EP}_{3}$. This further improves the dispersive readout method~\cite{Hodaei17}.

The proposed dispersive readout scheme only involves the weak-coupling regime~\cite{Xiang13,Kurizki15}, so it can suppress the unwanted backaction of the measurement on the qubit, as compared with the conventional dispersive readout method in the strong-coupling regime. Also, different from the proposal in Ref.~\cite{Troiani18}, the auxiliary cavities are spatially separated from the probed qubit to avoid the direct coupling between the auxiliary cavities and the qubit. Moreover, in contrast to the proposal in Ref.~\cite{Troiani18}, the effect of the qubit's decay is considered here, because it actually cannot be ignored in the weak-coupling regime. Combining the intriguing properties of the $\mathcal{PT}$ symmetry~\cite{Bender07,Konotop16} and the EPs~\cite{Lin11,Feng13,Lv15,Liertzer12,Feng14,Hodaei14,Lin16}, it is expected to explore more novel phenomena in the future by enhancing the sensitivity and precision of the quantum metrology.

\section{The model}\label{model}

\begin{figure}
\includegraphics[width=0.48\textwidth]{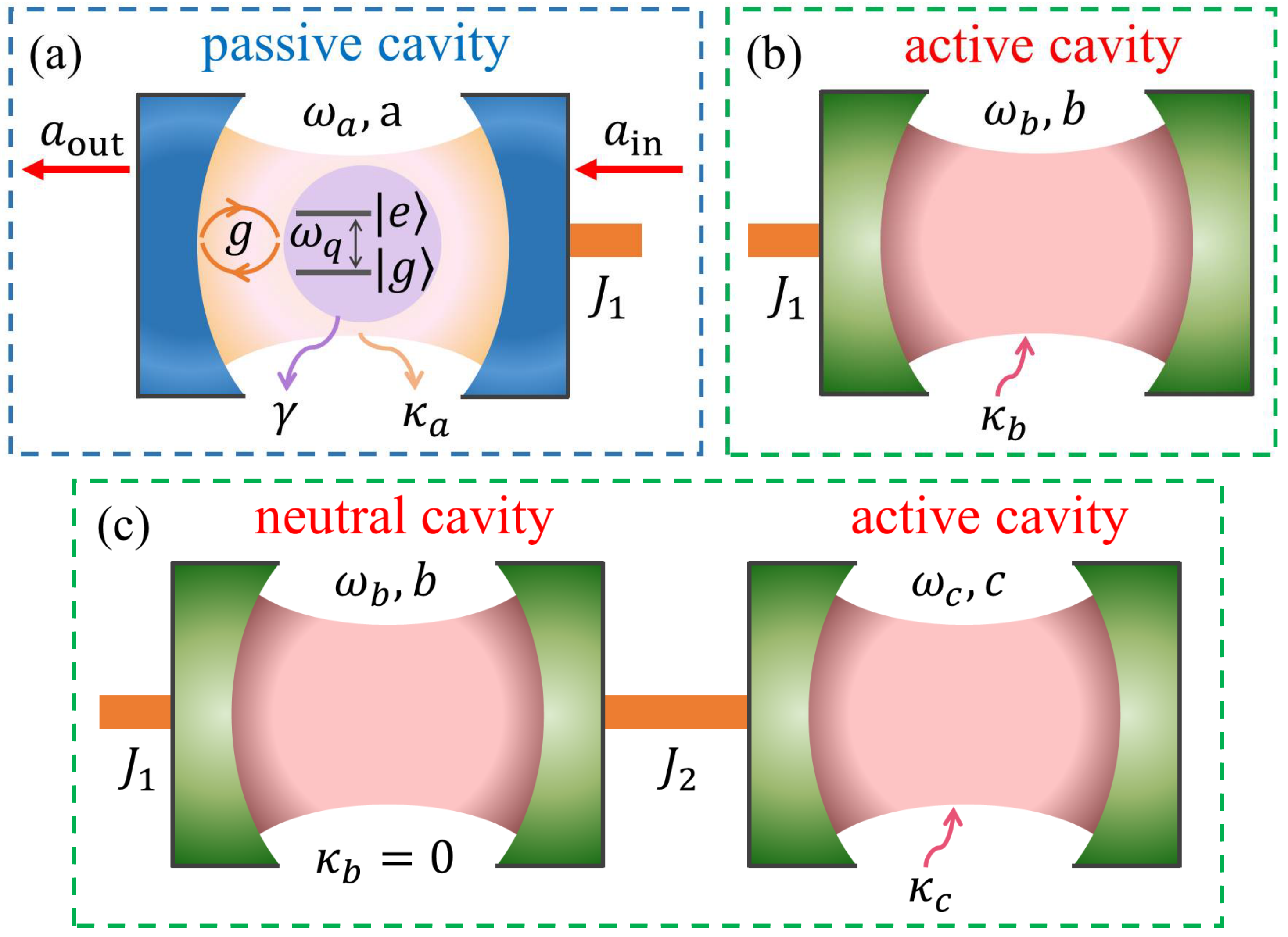}
\caption{Schematic of the proposed setup for the dispersive readout of (a) a qubit weakly coupled to a passive cavity, using either (b) an auxiliary cavity or (c) two auxiliary cavities linearly coupled to the passive cavity. Here, $a_{\rm in}$ and $a_{\rm out}$ are the input- and output-field operators.}
\label{fig1}
\end{figure}

For a qubit coupled to a cavity, the system is governed by the following Hamiltonian (setting $\hbar=1$):
\begin{equation}\label{equ1}
H_{s}=\omega_{a}a^{\dag}a+\frac{1}{2}\omega_{q}\sigma_{z}+g(a^{\dag}\sigma^{-}+a\sigma^{+}),
\end{equation}
where $a^{\dag}$ ($a$) is the creation (annihilation) operator of the cavity mode with frequency $\omega_{a}$, $\omega_{q}$ is the transition frequency of the qubit, $\sigma_{x}$, $\sigma_{y}$, and $\sigma_{z}$ are spin-1/2 Pauli operators, $\sigma^{\pm}=(\sigma_{x} \pm i\sigma_{y})/2$ are the ladder operators of the qubit, and $g$ is the coupling strength between the qubit and the cavity. In the dispersive regime, i.e., $g \ll \Delta_{q}$, with $\Delta_{q} \equiv \omega_{q}-\omega_{a}$ being the frequency detuning between the qubit and the cavity, the qubit is approximately decoupled from the cavity, but the cavity frequency is shifted from $\omega_a$ by $\pm g^{2}/\Delta_{q}$~\cite{Wallraff04}, where it is supposed that $\omega_{q} > \omega_{a}$, and $+$ ($-$) corresponds to the qubit being in the excited (ground) state. For a lossy (i.e., passive) cavity, as illustrated in Fig.~\ref{fig1}(a),
when the qubit is strongly coupled to the cavity (i.e., $g \gg \{\kappa_{a},~\gamma\}_{\rm max}$), the qubit state can be probed by measuring the transmission spectrum of the cavity $a$ because $g^{2}/\Delta_{q} > \kappa_{a}$, where $\kappa_{a}$ is the damping rate of the cavity mode and $\gamma$ is the decay rate of the qubit. However, in the weak-coupling regime (i.e., $g < \{\kappa_{a},~\gamma\}_{\rm min}$), the conventional dispersive readout scheme fails since $g^{2}/\Delta_{q} \ll \kappa_{a}$.

To measure the state of a weakly coupled qubit, one can couple the passive cavity to an auxiliary cavity with gain (i.e., an active cavity)~\cite{Quijandria18}; see Fig.~\ref{fig1}(b). The Hamiltonian of the auxiliary cavity is
\begin{equation}\label{equ2}
H_{\rm aux}=\omega_{b}b^{\dag}b,
\end{equation}
and the interaction Hamiltonian between the active and passive cavities is
\begin{equation}\label{equ3}
H_{\rm int}=J_{1}(a^{\dag}b+ab^{\dag}),
\end{equation}
where $b$ ($b^{\dag}$) is the annihilation (creation) operator of the auxiliary cavity mode with frequency $\omega_{b}$, and $J_{1}$ is the coupling strength between the active and passive cavities. The total Hamiltonian $H=H_{s}+H_{\rm aux}+H_{\rm int}$ of the system can be written as
\begin{equation}\label{equ4}
\begin{split}
H=&\;\omega_{a}a^{\dag}a+\omega_{b}b^{\dag}b+\frac{1}{2}\omega_{q}\sigma_{z}+g(a^{\dag}\sigma^{-}+a\sigma^{+})\\
  &+J_{1}(a^{\dag}b+ab^{\dag}).
\end{split}
\end{equation}
Eliminating the degree of freedom of the qubit via the quantum Langevin approach~\cite{Walls94}, we obtain the effective non-Hermitian Hamiltonian of the system (see Appendix~\ref{appendix-A1}),
\begin{equation}\label{equ5}
H_{\rm eff}^{(j)}=(\delta_{q}^{(j)}-i\kappa_{a})a^{\dag}a+(\Delta_{b}+i\kappa_{b})b^{\dag}b+J_{1}(a^{\dag}b+ab^{\dag}),
\end{equation}
where $\Delta_{b}=\omega_{b}-\omega_{a}$ is the frequency detuning between the passive and active cavities, $\kappa_{b}$ is the gain rate of the auxiliary active cavity $b$, and
\begin{equation}\label{equ6}
\begin{split}
&\delta_{q}^{(j)} \equiv \delta_{q}^{(e)}=+\frac{g^{2}}{\Delta_{q}-i\gamma},~~~j=e,\\
&\delta_{q}^{(j)} \equiv \delta_{q}^{(g)}=-\frac{g^{2}}{\Delta_{q}-i\gamma},~~~j=g,
\end{split}
\end{equation}
are the qubit-induced frequency shifts of the lossy cavity mode when the qubit is in the excited and ground states (i.e., $j=e$ and $j=g$), respectively. Under the dispersive strong-coupling condition $\Delta_{q} \gg g \gg  \{\kappa_{a},~\gamma\}_{\rm max}$, $\delta_{q}^{(j)}$ is reduced to $\delta_{q}^{(j)} = \pm g^{2}/\Delta_{q}$, with $\Delta_{q}-i\gamma \approx \Delta_{q}$. However, the decay rate of the qubit cannot be ignored~\cite{Troiani18} in the dispersive weak-coupling regime, $\Delta_{q} \gg g$ and $g < \{\kappa_{a},~\gamma\}_{\rm min}$, because the relation $\Delta_{q} \gg \gamma$ becomes invalid. In this case, we can treat the weakly coupled qubit as a perturbation acting on the passive cavity, since $|\delta_{q}^{(j)}| \ll \kappa_{a}$.

In order to further improve the dispersive readout method~\cite{Hodaei17}, we introduce another auxiliary cavity $c$ coupled to the first auxiliary cavity $b$ [see Fig.~\ref{fig1}(c)]. The interaction Hamiltonian $H_{\rm int}$ is the same as in Eq.~(\ref{equ3}), but the Hamiltonian $H_{\rm aux}$ of the double auxiliary cavities is
\begin{equation}\label{equ7}
H_{\rm aux}=\omega_{b}b^{\dag}b+\omega_{c}c^{\dag}c+J_{2}(b^{\dag}c+bc^{\dag}),
\end{equation}
where $c$ ($c^{\dag}$) is the annihilation (creation) operator of the cavity mode with frequency $\omega_{c}$ in the second auxiliary cavity, and $J_{2}$ is the coupling strength between the two auxiliary cavities. The total Hamiltonian $H=H_{s}+H_{\rm aux}+H_{\rm int}$ of the system is now given by
\begin{equation}\label{equ8}
\begin{split}
H=\;&\omega_{a}a^{\dag}a+\omega_{b}b^{\dag}b+\omega_{c}c^{\dag}c+\frac{1}{2}\omega_{q}\sigma_{z}+g(a^{\dag}\sigma^{-}+a\sigma^{+})\\
  &+J_{1}(a^{\dag}b+ab^{\dag})+J_{2}(b^{\dag}c+bc^{\dag}).
\end{split}
\end{equation}
Also, eliminating the degree of freedom of the qubit and including both loss and gain in the system, we obtain the effective non-Hermitian Hamiltonian of the system (see Appendix~\ref{appendix-A2})
\begin{equation}\label{equ9}
\begin{split}
H_{\rm eff}^{(j)}=\;&(\delta_{q}^{(j)}-i\kappa_{a})a^{\dag}a+(\Delta_{b}+i\kappa_{b})b^{\dag}b+(\Delta_{c}+i\kappa_{c})c^{\dag}c\\
                  & +J_{1}(a^{\dag}b+ab^{\dag})+J_{2}(b^{\dag}c+bc^{\dag}),
\end{split}
\end{equation}
where $\Delta_{c}=\omega_{c}-\omega_{a}$ is the frequency detuning of the cavity $c$ from the cavity $a$, and $\kappa_{c}$ is the gain rate of the cavity $c$.

\section{Readout of a qubit around the EP$_{2}$}\label{one}

\begin{figure}
\includegraphics[width=0.45\textwidth]{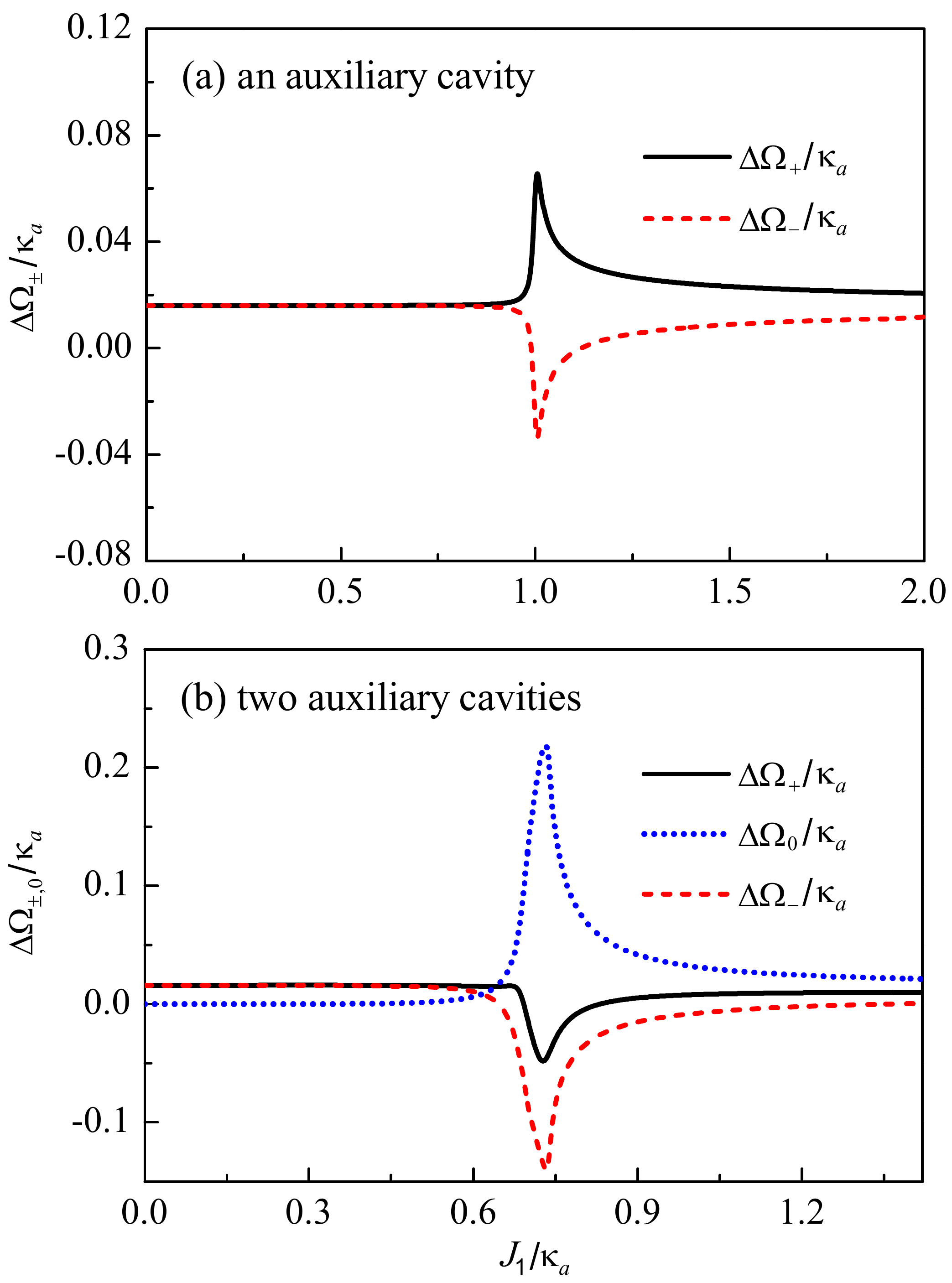}
\caption{(a) The difference $\Delta\Omega_{\pm}/\kappa_{a}$ between the real parts of the eigenfrequencies $\Omega_{\pm}^{(g)}$ and $\Omega_{\pm}^{(e)}$, ${\rm Re}[\Omega_{\pm}^{(g)}]$ and ${\rm Re}[\Omega_{\pm}^{(e)}]$, vs the coupling strength $J_{1}/\kappa_{a}$ in the case of an auxiliary cavity with $\kappa_{b}/\kappa_{a}=1$. (b) The difference $\Delta\Omega_{\pm,0}/\kappa_{a}$ between the real parts of the eigenfrequencies $\Omega_{\pm,0}^{(g)}$ and $\Omega_{\pm,0}^{(e)}$, ${\rm Re}[\Omega_{\pm,0}^{(g)}]$ and ${\rm Re}[\Omega_{\pm,0}^{(e)}]$, vs the coupling strength $J_{1}/\kappa_{a}$ in the case of two auxiliary cavities with $\Delta_{c}=\kappa_{b}=0$, $J_{1}=J_{2}$, and $\kappa_{c}/\kappa_{a}=1$. Other parameters are chosen to be $\Delta_{b}=0$, $\gamma/\kappa_{a}=1$, $g/\kappa_{a}=0.2$, and $\Delta_{q}=10g$.}
\label{fig2}
\end{figure}

Without the qubit, i.e., $\delta_{q}^{(j)}=0$, the effective Hamiltonian $H_{\rm eff}^{(j)}$ in Eq.~(\ref{equ5}) is reduced to a Hamiltonian with the $\mathcal{PT}$ symmetry,
\begin{equation}\label{equ10}
H_{\rm PT} =-i\kappa_{a}a^{\dag}a+i\kappa_{a}b^{\dag}b+J_{1}(a^{\dag}b+ab^{\dag}),
\end{equation}
when the passive cavity is resonant with the auxiliary cavity ($\Delta_{b}=0$) and both the gain and loss of the two cavities are balanced ($\kappa_{b}/\kappa_{a}=1$). Diagonalizing the $\mathcal{PT}$-symmetric Hamiltonian $H_{\rm PT}$, we obtain the two eigenfrequencies of the system's supermodes $d_{\pm}$,
\begin{equation}\label{equ11}
\Omega_{\pm}=\pm\sqrt{J_{1}^{2}-\kappa_{a}^{2}}.
\end{equation}
The system can experience a phase transition from the $\mathcal{PT}$-symmetric phase with real $\Omega_{\pm}$ to the $\mathcal{PT}$-symmetry-breaking phase with complex $\Omega_{\pm}$, when $J_{1}/\kappa_{a}$ varies from $J_{1}/\kappa_{a} > 1$ to $J_{1}/\kappa_{a} < 1$. For a non-Hermitian system, the real and imaginary parts of the eigenvalues $\Omega_{\pm}$, ${\rm Re}[\Omega_{\pm}]$ and ${\rm Im}[\Omega_{\pm}]$, represent the frequency detunings of the supermodes $d_{\pm}$ from the cavity mode $a$ and their loss or gain rates, respectively. At the critical point (i.e., the EP$_{2}$) with $J_{\rm EP2}/\kappa_{a}=1$, the two eigenvalues coalesce to $\Omega_{\pm}=0$. It is at the EP that the detection sensitivity of the frequency or energy splitting can be enhanced~\cite{Wiersig14,Liu16,Hodaei17,Chen17}. Thus, one can probe the state of a weakly coupled qubit around the EP$_{2}$ by measuring the transmission spectrum of the lossy cavity.

When including the qubit, the two eigenfrequencies $\Omega_{\pm}$ of the system's supermodes $d_{\pm}$ become
\begin{equation}\label{equ12}
\Omega_{\pm}^{(j)}=\frac{\delta_{q}^{(j)}}{2}
                    \pm \sqrt{J_{1}^{2}+\Bigg(\frac{\delta_{q}^{(j)}}{2}\Bigg)^{2}-\kappa_{a}^{2}-i\kappa_{a}\delta_{q}^{(j)}},
\end{equation}
as obtained by diagonalizing the effective Hamiltonian $H_{\rm eff}^{(j)}$ in Eq.~(\ref{equ5}) under the $\mathcal{PT}$-symmetric condition. Note that the parameter in the square root of Eq.~(\ref{equ12}) becomes complex when the qubit is included in the Hamiltonian of the system. For a complex number $z=\rho e^{i\theta}$, where $\rho$ and $\theta$ are real, its square root has two values $\zeta=\sqrt{z}=\sqrt{\rho} e^{i(\theta+2k\pi)/2}=\pm\sqrt{\rho}e^{i\theta/2}$, with $k=0$, $1$. Obviously, there is no effect on the eigenvalues of the system when choosing either $k=0$ or 1 for Eq.~(\ref{equ12}). Hereafter, we choose $k=0$ and $-\pi < \theta \leq \pi$ in our numerical simulation. In the transmission spectrum of the passive cavity $a$, there are two peaks corresponding to the two eigenfrequencies $\Omega_{\pm}^{(j)}$ for $j=e$ or $g$, where ${\rm Re}[\Omega_{\pm}^{(j)}]$ and ${\rm Im}[\Omega_{\pm}^{(j)}]$ determine the locations and linewidths of the two peaks, respectively~\cite{Kurucz11}.

To show that the perturbation induced by the qubit can be amplified by the EP$_2$, we introduce two experimentally measurable quantities,
\begin{equation}\label{equ13}
\Delta\Omega_{\pm}={\rm Re}[\Omega_{\pm}^{(e)}]-{\rm Re}[\Omega_{\pm}^{(g)}],
\end{equation}
which represent the frequency differences of the system when the qubit is in the ground and excited states, respectively. In Fig.~\ref{fig2}(a), we display the differences $\Delta\Omega_{\pm}/\kappa_{a}$ versus the coupling strength $J_{1}/\kappa_{a}$. When $J_{1}/\kappa_{a}$ varies from 0 (i.e., without the auxiliary cavity) to $\sim 1$ (i.e., around the EP$_{2}$), the difference $\Delta\Omega_{+}/\kappa_{a}$ increases from 0.016 to 0.066 (solid black curve) and $\Delta\Omega_{-}/\kappa_{a}$ decreases to -0.034 (dashed red curve). However, $\Delta\Omega_{+}/\kappa_{a}$ sharply decreases to 0.020, and $\Delta\Omega_{-}/\kappa_{a}$ increases to 0.011 as the coupling strength $J_{1}/\kappa_{a}$ further increases to 2. The peak (dip) around $J_{1}=J_{\rm EP2}$ corresponds to the maximum (minimum) value of $\Delta\Omega_{+}/\kappa_{a}$ ($\Delta\Omega_{-}/\kappa_{a}$), which can be used to resolve the states of the weakly coupled qubit.

\begin{figure}
\includegraphics[width=0.48\textwidth]{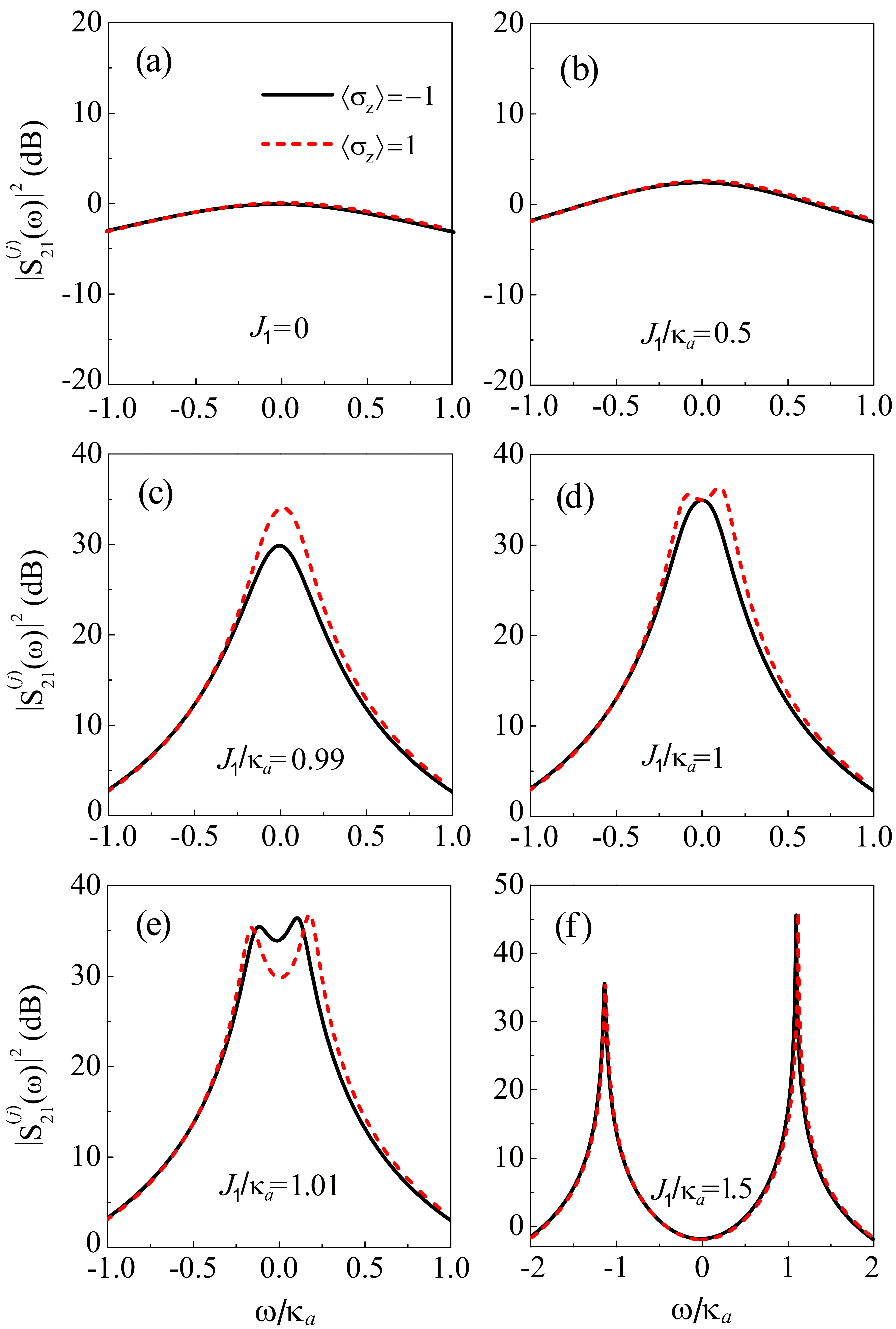}
\caption{The transmission spectrum $|S_{21}^{(j)}(\omega)|^{2}$ of the passive cavity vs the probe-field frequency $\omega/\kappa_{a}$ in the case of an auxiliary cavity, where $\kappa_{i}/\kappa_{a}=\kappa_{o}/\kappa_{a}=1/2$ and (a) $J_{1}=0$, (b) $J_{1}/\kappa_{a}=0.5$, (c) $J_{1}/\kappa_{a}=0.99$, (d) $J_{1}/\kappa_{a}=1$, (e) $J_{1}/\kappa_{a}=1.01$, and (f) $J_{1}/\kappa_{a}=1.5$. Other parameters are the same as in Fig.~\ref{fig2}(a).}
\label{fig3}
\end{figure}

In the experiment, the dispersive readout of the qubit can be realized by measuring the transmission spectrum. As shown in Fig.~\ref{fig1}(a), there is an input (output) field $a_{\rm in}$ ($a_{\rm out}$) with frequency $\omega$ acting on the input (output) port of the cavity $a$. With the relation $a_{\rm out}^{(j)}=S_{21}^{(j)}(\omega)a_{\rm in}$, we can define the transmission coefficient $S_{21}^{(j)}(\omega)$ of the passive cavity (see Appendix~\ref{appendix-B}),
\begin{equation}\label{equ15}
S_{21}^{(j)}(\omega)=\frac{2\sqrt{\kappa_{i}\kappa_{o}}}{\kappa_{a}+i(\delta_{q}^{(j)}-\omega)+\sum(\omega)},
\end{equation}
where the decay rate $\kappa_{i}$ ($\kappa_{o}$) of the passive cavity $a$ is induced by the input (output) port, and
\begin{equation}\label{equ16}
\sum(\omega)=\frac{J_{1}^{2}}{-\kappa_{b}+i(\Delta_{b}-\omega)}
\end{equation}
is the self-energy resulting from cavity $b$. If the transmission spectra $S_{21}^{(g)}(\omega)$ and $S_{21}^{(e)}(\omega)$ have a clear difference, the qubit state can then be resolved by measuring the transmission spectrum $S_{21}^{(j)}(\omega)$ of the passive cavity.

In the absence of the auxiliary cavity (i.e., $J_{1}=0$), the states of the qubit weakly coupled to a passive cavity cannot be resolved when using the dispersive readout method [see Fig.~\ref{fig3}(a)].
For $J_{1}/\kappa_{a}\neq 0$ but away from the EP$_{2}$, it is still difficult to resolve the quantum states of the weakly coupled qubit [see Figs.~\ref{fig3}(b) and \ref{fig3}(f) with $J_{1}/\kappa_{a}=0.5$ and $1.5$, respectively].
However, an appreciable difference between the transmission spectra $S_{21}^{(g)}(\omega)$ and $S_{21}^{(e)}(\omega)$ occurs when the coupling strength $J_{1}/\kappa_{a}$ between the active and passive cavities approaches the critical value $J_{1}=J_{\rm EP2}$ [see Figs.~\ref{fig3}(c)-\ref{fig3}(e)]. There are two reasons giving rise to this phenomenon: (i) the difference $\Delta \Omega_{\pm}$ of the energy spectrum is amplified near the EP$_{2}$ [see Fig.~\ref{fig2}(a)] and (ii) the $\mathcal{PT}$ symmetry of the system narrows the linewidth of the peak in the transmission spectrum of the passive cavity. Therefore, one can measure the states of the weakly coupled qubit using an auxiliary cavity around the EP$_{2}$.

\begin{figure}
\includegraphics[width=0.48\textwidth]{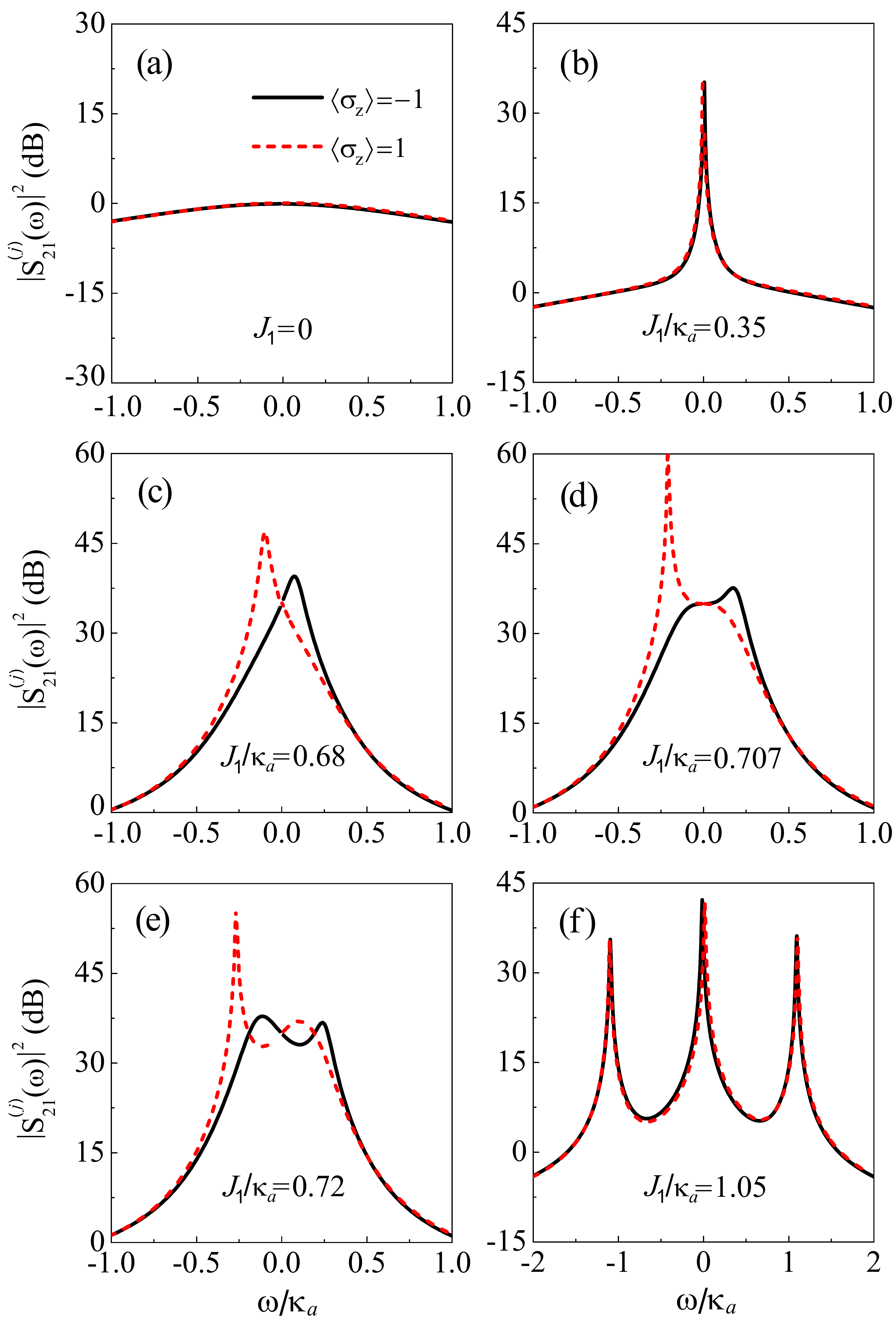}
\caption{The transmission spectrum $|S_{21}^{(j)}(\omega)|^{2}$ of the passive cavity vs the probe-field frequency $\omega/\kappa_{a}$ in the case of two auxiliary cavities, where $\kappa_{i}/\kappa_{a}=\kappa_{o}/\kappa_{a}=1/2$ and (a) $J_{1}=0$, (b) $J_{1}/\kappa_{a}=0.35$, (c) $J_{1}/\kappa_{a}=0.68$, (d) $J_{1}/\kappa_{a}=0.707$, (e) $J_{1}/\kappa_{a}=0.72$, and (f) $J_{1}/\kappa_{a}=1.05$. Other parameters are the same as in Fig.~\ref{fig2}(b).}
\label{fig4}
\end{figure}

\section{Readout of a qubit around the EP$_{3}$}\label{double}

For the case of two auxiliary cavities, the $\mathcal{PT}$-symmetric condition becomes $\Delta_{b}=\Delta_{c}=0$, $J_{1}=J_{2}$, $\kappa_{b}=0$, and $\kappa_{c}/\kappa_{a}=1$ when excluding the qubit. Under this condition, the effective non-Hermitian Hamiltonian $H_{\rm eff}^{(j)}$ in Eq.~(\ref{equ9}) is reduced to a $\mathcal{PT}$-symmetric Hamiltonian,
\begin{equation}\label{equ18}
\begin{split}
H_{\rm PT} = & -i\kappa_{a}a^{\dag}a+i\kappa_{a}c^{\dag}c +J_{1}(a^{\dag}b+ab^{\dag})+J_{1}(b^{\dag}c+bc^{\dag}),
\end{split}
\end{equation}
which has three eigenvalues,
\begin{equation}\label{equ19}
\Omega_{\pm}=\pm\sqrt{2J_{1}^{2}-\kappa_{a}^{2}},~~~\Omega_{0}=0.
\end{equation}
When $2J_{1}^{2}-\kappa_{a}^{2} = 0$ (i.e., $J_{\rm EP3}/\kappa_{a}=\sqrt{2}/2$), the three eigenvalues coalesce to the EP$_{3}$, $\Omega_{\pm}=\Omega_{0}$. Obviously, the two eigenvalues $\Omega_{\pm}$ are real (complex) for $J_{1} > J_{\rm EP3}$ ($J_{1} < J_{\rm EP3} $).

Different from the case of an auxiliary cavity, we do not have analytical expressions for the eigenvalues of the system, $\Omega_{\pm}^{(j)}$ and $\Omega_{0}^{(j)}$, when including the qubit, where $j=e$ ($j=g$) corresponds to the qubit being in the excited (ground) state. In this case of two auxiliary cavities, we can also define the differences $\Delta\Omega_{\pm,0}$ between the real parts of the eigenfrequencies $\Omega_{\pm,0}^{(g)}$ and $\Omega_{\pm,0}^{(e)}$, with $\Delta\Omega_{\pm}$ in the same form as Eq.~(\ref{equ13}), and $\Delta\Omega_{0}={\rm Re}[\Omega_{0}^{(e)}]-{\rm Re}[\Omega_{0}^{(g)}]$. Figure~\ref{fig2}(b) displays $\Delta\Omega_{\pm,0}$ versus the coupling strength $J_{1}/\kappa_{a}$ in the case of two auxiliary cavities. When $J_{1} \approx J_{\rm EP3}$, the difference $\Delta\Omega_{0}$ reaches its peak and $\Delta\Omega_{\pm}$ approach their dips, respectively, near which it is appropriate to distinguish the qubit states. Comparing Fig.~\ref{fig2}(b) with Fig.~\ref{fig2}(a), we see that the perturbation induced by the qubit (i.e., the differences between the real parts of the system's eigenfrequencies for the qubit being in the ground and excited states) can be further amplified near the EP$_{3}$.

In the considered case of two auxiliary cavities, the transmission spectrum $S_{21}^{(j)}(\omega)$ of the passive cavity in Eq.~(\ref{equ15}) is still valid, but the corresponding self-energy in Eq.~(\ref{equ16}) becomes (see Appendix~\ref{appendix-B})
\begin{equation}\label{equ21}
\sum(\omega)=\frac{J_{1}^{2}}{-\kappa_{b}+i(\Delta_{b}-\omega)+J_{2}^{2}/[-\kappa_{c}+i(\Delta_{c}-\omega)]}.
\end{equation}
In Fig.~\ref{fig4}, we plot the transmission spectrum $|S_{21}^{(j)}(\omega)|^{2}$ of the passive cavity versus the probe-field frequency $\omega/\kappa_{a}$ for various values of the coupling strength $J_{1}/\kappa_{a}$. Different from the case without auxiliary cavities [see Fig.~\ref{fig4}(a)], i.e., $J_{1}=0$, $|S_{21}^{(g)}(\omega)|^{2}$ and $|S_{21}^{(e)}(\omega)|^{2}$ can be easily distinguished from each other around $J_{1}=J_{\rm EP3}$ [see Figs.~\ref{fig4}(c)-\ref{fig4}(e)].
Similar to the case of an auxiliary cavity, $|S_{21}^{(g)}(\omega)|^{2}$ and $|S_{21}^{(e)}(\omega)|^{2}$ are indistinguishable when the system's parameters are away from the EP$_{3}$ conditions [see Figs.~\ref{fig4}(b) and \ref{fig4}(f) with $J_{1}/\kappa_{a}=0.35$ and $1.05$, respectively]. In comparison with the case of a single auxiliary cavity, the results also verify that the scheme with two auxiliary cavities can further improve the sensitivity when measuring the states of the weakly coupled qubit, because the difference between the transmission spectra $S_{21}^{(g)}(\omega)$ and $S_{21}^{(e)}(\omega)$ becomes more appreciable [cf.~Figs.~\ref{fig3}(c)-\ref{fig3}(e) and Figs.~\ref{fig4}(c)-\ref{fig4}(e)].

\section{Discussion and conclusion}\label{discussion}

For a 2D SC cavity, there may exist a pair of modes with close frequencies, only one of which is weakly coupled to a qubit~\cite{Troiani18}. In that case, the qubit states can be dispersively measured by using the other mode as an auxiliary mode. However, such a scheme has some limitations due to the inevitable coupling between the qubit and the auxiliary mode~\cite{Bonizzoni18} as well as the neglect of the qubit's decay. In our proposal, the auxiliary component is either the one or the two 1D SC cavities, which can be spatially separated from the qubit to avoid the direct coupling between the auxiliary component and the qubit. Also, the decay rate of the weakly coupled qubit is included because it actually cannot be ignored in the weak-coupling regime. Moreover, as compared with the 2D SC cavity, the 1D SC cavity is commonly used in the circuit-QED experiment and has a larger rms zero-point cavity field due to its smaller effective volume (which can give rise to a stronger coupling strength).

In Ref.~\cite{Quijandria18}, two coupled SC cavities with $\mathcal{PT}$ symmetry, one with gain and the other with loss, have been theoretically proposed, where the gain is realized using an auxiliary SC qubit. In addition, a tunable coupling between two SC cavities has also been explored~\cite{Peropadre13,Baust15}. With these available conditions, our scheme can be experimentally implementable in SC circuits, where the weakly coupled qubit can be either an electron spin or a SC qubit. In a $\mathcal{PT}$-symmetric system, the spectral splitting resulting from a small perturbation $\xi$ follows a $\xi^{1/n}$ dependence near an EP$_{n}$~\cite{Hodaei17,Chen17}. Theoretically, one can increase the eigenfrequency difference of the system for the qubit being in the ground and excited states by harnessing even more auxiliary cavities to synthesize a higher-order EP. However, it becomes an experimental challenge when synthesizing such a higher-order EP because more system's parameters should be tuned simultaneously to satisfy the conditions of an EP$_{n}$.

For available experimental parameters, some non-Hermitian quantum systems are implemented probabilistically (see, e.g., Refs.~\cite{Lee14,Scheel18,Kawabata17}). Specifically, in the one-dimensional non-Hermitian \emph{XY} model studied in Ref.~\cite{Lee14}, the non-Hermiticity is from measuring whether or not a spontaneous decay has occurred in an atom, which is probabilistic. However, in our approach, the non-Hermitian systems are implemented in a deterministic manner, with given losses and gains related to the non-Hermiticity of the systems.

In summary, we have proposed two schemes to measure the quantum states of a qubit weakly coupled to a passive cavity in the dispersive regime. For a circuit QED in the weak-coupling regime~\cite{Xiang13,Kurizki15}, it is difficult to measure the states of a qubit with the conventional dispersive readout method. However, in our scheme, we employ either an auxiliary cavity or two auxiliary cavities to form an EP$_{2}$ or EP$_{3}$ when the weakly coupled qubit is ignored. By studying the energy spectrum, we find that the difference of the energy spectra for the qubit being in the ground and excited states, respectively, can be amplified near the EP$_{2}$ and EP$_{3}$~\cite{Wiersig14,Liu16,Hodaei17,Chen17}, which is measurable by probing the transmission spectrum of the passive cavity in the experiment. Our improved dispersive readout method paves a way to measure the quantum states of a qubit weakly coupled to a passive cavity. Compared with the conventional dispersive readout method in the strong-coupling regime, our schemes can also reduce the backaction induced by the measurement on the weakly coupled qubit.

\section*{Acknowledgments}

This work is supported by the National Key Research and Development Program of China (Grant No.~2016YFA0301200) and the National Natural Science Foundation of China (Grants No.~U1801661 and No.~11774022).

\appendix

\section{The effective non-Hermitian Hamiltonian}\label{appendix-A}

\subsection{The case of an auxiliary cavity}\label{appendix-A1}

In a rotating frame with respect to the frequency $\omega_{a}$ of the passive cavity, the total Hamiltonian $H$ in Eq.~(\ref{equ4}) becomes
\begin{equation}\label{a1}
H=\Delta_{b}b^{\dag}b+\frac{1}{2}\Delta_{q}\sigma_{z}+g(a^{\dag}\sigma^{-}+a\sigma^{+})+J_{1}(a^{\dag}b+ab^{\dag}),
\end{equation}
where $\Delta_{b}=\omega_{b}-\omega_{a}$ is the frequency detuning between cavity modes $a$ and $b$. When a probe field $a_{\rm in}$ (i.e., the input field) acts on the passive cavity, the quantum dynamics of the system is governed by the following quantum Langevin equations~\cite{Walls94}:
\begin{equation}\label{a2}
\begin{split}
&\dot{a}=-\kappa_{a}a-ig\sigma^{-}-iJ_{1}b+\sqrt{2\kappa_{i}}\,a_{\rm in},\\
&\dot{b}=-i(\Delta_{b}+i\kappa_{b})b-iJ_{1}a,\\
&\dot{\sigma}^{-}=-i(\Delta_{q}-i\gamma)\sigma^{-}+iga\sigma_{z},
\end{split}
\end{equation}
where $\kappa_{b}$ is the gain rate of the auxiliary cavity and $\kappa_{i}$ denotes the coupling strength between the probe field $a_{\rm in}$ and the mode $a$ of the lossy cavity. In the dispersive regime $g \ll \Delta_{q}$, we suppose that the qubit is in the steady state. Then, we obtain
\begin{equation}\label{a3}
\sigma^{-}=\frac{g}{\Delta_{q}-i\gamma}a\sigma_{z},
\end{equation}
by solving the third equation in Eq.~(\ref{a2}) with $\dot{\sigma}^{-}=0$.

Substituting the above expression of $\sigma^{-}$ into the first equation in Eq.~(\ref{a2}) and then replacing the operator $\sigma_{z}$ with its eigenvalues $\pm 1$ to eliminate the degree of freedom of the qubit, we have
\begin{equation}\label{a4}
\begin{split}
&\dot{a}=-i(\delta_{q}^{(j)}-i\kappa_{a})a-iJ_{1}b+\sqrt{2\kappa_{i}}\,a_{\rm in} ,\\
&\dot{b}=-i(\Delta_{b}+i\kappa_{b})b-iJ_{1}a,
\end{split}
\end{equation}
where the qubit-induced frequency shifts $\delta_{q}^{(j)}$ ($j=e,~g$) of the passive cavity mode are given in Eq.~(\ref{equ6}). The quantum Langevin equations in Eq.~(\ref{a4}) can be rewritten as
\begin{equation}\label{a5}
\dot{a}=-i[a,H_{\rm eff}^{(j)}]+\sqrt{2\kappa_{i}}\,a_{\rm in},~~~~\dot{b}=-i[b,H_{\rm eff}^{(j)}],
\end{equation}
where $H_{\rm eff}^{(j)}$, as given in Eq.~(\ref{equ5}), is the effective non-Hermitian Hamiltonian of the system.
\subsection{The case of two auxiliary cavities}\label{appendix-A2}

Also in a rotating frame with respect to the frequency $\omega_{a}$ of the passive cavity, we can write
the total Hamiltonian in Eq.~(\ref{equ8}) as
\begin{equation}\label{a6}
\begin{split}
H=\;&\Delta_{b}b^{\dag}b+\Delta_{c}c^{\dag}c+\frac{1}{2}\Delta_{q}\sigma_{z}+g(a^{\dag}\sigma^{-}+a\sigma^{+})\\
  &+J_{1}(a^{\dag}b+ab^{\dag})+J_{2}(b^{\dag}c+bc^{\dag}),
\end{split}
\end{equation}
where $\Delta_{c}=\omega_{c}-\omega_{a}$ is the frequency detuning between cavity modes $a$ and $c$. Using the quantum  Langevin approach~\cite{Walls94}, we can obtain the following equations to describe the quantum dynamics of the system:
\begin{equation}\label{a7}
\begin{split}
&\dot{a}=-\kappa_{a}a-ig\sigma^{-}-iJ_{1}b+\sqrt{2\kappa_{i}}\,a_{\rm in},\\
&\dot{b}=-i(\Delta_{b}+i\kappa_{b})b-iJ_{1}a-iJ_{2}c,\\
&\dot{c}=-i(\Delta_{c}+i\kappa_{c})c-iJ_{2}b,\\
&\dot{\sigma}^{-}=-i(\Delta_{q}-i\gamma)\sigma^{-}+iga\sigma_{z},
\end{split}
\end{equation}
with $\kappa_{c}$ being the gain rate of the cavity $c$. Assuming the qubit is in the steady state and eliminating its degree of freedom, we have
\begin{equation}\label{a8}
\begin{split}
&\dot{a}=-i(\delta_{q}^{(j)}-i\kappa_{a})a-iJ_{1}b+\sqrt{2\kappa_{i}}\,a_{\rm in} ,\\
&\dot{b}=-i(\Delta_{b}+i\kappa_{b})b-iJ_{1}a-iJ_{2}c,\\
&\dot{c}=-i(\Delta_{c}+i\kappa_{c})c-iJ_{2}b.
\end{split}
\end{equation}
We can convert the first two equations in Eq.~(\ref{a8}) to the same compact form as in Eq.~(\ref{a5}) and rewrite the third equation in Eq.~(\ref{a8}) as
\begin{equation}\label{a9}
\dot{c}=-i[c,H_{\rm eff}^{(j)}],
\end{equation}
where $H_{\rm eff}^{(j)}$ is the effective non-Hermitian Hamiltonian in Eq.~(\ref{equ9}).

\section{Transmission coefficient}\label{appendix-B}

In both cases of single and double auxiliary cavities, we can derive the transmission coefficient of cavity $a$. By performing the Fourier transform, the quantum Langevin equations in Eq.~(\ref{a4}) can be converted to
\begin{equation}\label{b1}
\begin{split}
&-i(\delta_{q}^{(j)}-\omega-i\kappa_{a})a-iJ_{1}b+\sqrt{2\kappa_{i}}\,a_{\rm in}=0,\\
&-i(\Delta_{b}-\omega+i\kappa_{b})b-iJ_{1}a=0,
\end{split}
\end{equation}
where $\omega$ is the frequency of the probe field $a_{\rm in}$ fed into the cavity via the input port. Solving the above equation, the field $a$ of the passive cavity can be expressed as
\begin{equation}\label{b2}
a^{(j)}=\frac{\sqrt{2\kappa_{i}}\,a_{\rm in}}{\kappa_{a}+i(\delta_{q}^{(j)}-\omega)+\sum(\omega)},
\end{equation}
where the self-energy $\sum(\omega)$ induced by the cavity with gain is given in Eq.~(\ref{equ16}). According to the input-output theory \cite{Walls94}, the intracavity field $a$ can be connected with the output field $a_{\rm out}$, which goes out from the passive cavity at the output port, via the relation
\begin{equation}\label{b3}
a_{\rm out}^{(j)}=\sqrt{2\kappa_{o}}\,a^{(j)},
\end{equation}
where $\kappa_{o}$ is the coupling strength between modes $a$ and $a_{\rm out}$. Here we have assumed that there is no input field at the output port. Combining Eqs.~(\ref{b2}) and (\ref{b3}) and using the relation $a_{\rm out}^{(j)}=S_{21}^{(j)}(\omega)a_{\rm in}$, we obtain the transmission coefficient $S_{21}^{(j)}(\omega)$ of the passive cavity given in Eq.~(\ref{equ15}).

For the case of two auxiliary cavities, via the Fourier transform, the quantum Langevin equations in Eq.~(\ref{a8}) can be converted to
\begin{equation}\label{b4}
\begin{split}
&-i(\delta_{q}^{(j)}-\omega-i\kappa_{a})a-iJ_{1}b+\sqrt{2\kappa_{i}}\,a_{\rm in}=0 ,\\
&-i(\Delta_{b}-\omega+i\kappa_{b})b-iJ_{1}a-iJ_{2}c=0,\\
&-i(\Delta_{c}-\omega+i\kappa_{c})c-iJ_{2}b=0.
\end{split}
\end{equation}
From Eq.~(\ref{b4}), it is easy to verify that the expression of operator $a$ can also be written in the same form as in Eq.~(\ref{b2}), but the corresponding self-energy is given by Eq.~(\ref{equ21}). With this self-energy, the transmission coefficient $S_{21}^{(j)}(\omega)$ of the passive cavity in Eq.~(\ref{equ15}) can then be obtained.


\begin{thebibliography}{99}

\bibitem{Nielsen00}
M. A. Nielsen and I. L. Chuang, \emph{Quantum Computation and Quantum Information} (Cambridge University Press, Cambridge, 2000).

\bibitem{Bennett00}
C. H. Bennett and D. P. DiVincenzo,
Quantum information and computation,
Nature (London) \textbf{404}, 247 (2000).

\bibitem{Steane98}
A. Steane,
Quantum Computing,
Rep. Prog. Phys. \textbf{61}, 117 (1998).


\bibitem{Schoelkopf08}
R. J. Schoelkopf and S. M. Girvin,
Wiring up quantum systems,
Nature (London) \textbf{451}, 664 (2008).

\bibitem{Xiang13}
Z. L. Xiang, S. Ashhab, J. Q. You, and F. Nori,
Hybrid quantum circuits: Superconducting circuits interacting with other quantum systems,
Rev. Mod. Phys. \textbf{85}, 623 (2013).

\bibitem{Kurizki15}
G. Kurizki, P. Bertet, Y. Kubo, K. M{\o}lmer, D. Petrosyan, P. Rabl, and J. Schmiedmayer,
Quantum technologies with hybrid systems,
Proc. Natl. Acad. Sci. U.S.A. {\bf 112}, 3866 (2015).

\bibitem{Wallraff04}
A. Wallraff, D. I. Schuster, A. Blais, L. Frunzio, R. S. Huang, J. Majer, S. Kumar, S. M. Girvin, and R. J. Schoelkopf,
Strong coupling of a single photon to a superconducting qubit using circuit quantum electrodynamics,
Nature (London) \textbf{431}, 162 (2004).

\bibitem{Kohler17}
S. Kohler,
Dispersive Readout of Adiabatic Phases,
Phys. Rev. Lett. \textbf{119}, 196802 (2017).

\bibitem{Haigh15}
J. A. Haigh, N. J. Lambert, A. C. Doherty, and A. J. Ferguson,
Dispersive readout of ferromagnetic resonance for strongly coupled magnons and microwave photons,
Phys. Rev. B \textbf{91}, 104410 (2015).

\bibitem{Benito16}
M. Benito, X. Mi, J. M. Taylor, J. R. Petta, and G. Burkard,
Input-output theory for spin-photon coupling in Si double quantum dots,
Phys. Rev. B \textbf{96}, 235434 (2017).

\bibitem{Burkard16}
G. Burkard and J. R. Petta,
Dispersive readout of valley splittings in cavity-coupled silicon quantum dots,
Phys. Rev. B \textbf{94}, 195305 (2016).

\bibitem{Zueco09}
D. Zueco, G. M. Reuther, S. Kohler, and P. H\"{a}nggi,
Qubit-oscillator dynamics in the dispersive regime: Analytical theory beyond the rotating-wave approximation,
Phys. Rev. A \textbf{80}, 033846 (2009).

\bibitem{Kohler18}
S. Kohler,
Dispersive readout: Universal theory beyond the rotating-wave approximation,
Phys. Rev. A \textbf{98}, 023849 (2018).

\bibitem{Niemczyk10}
T. Niemczyk, F. Deppe, H. Huebl, E. P. Menzel, F. Hocke, M. J. Schwarz, J. J. Garcia-Ripoll, D. Zueco, T. H\"{u}mmer, E. Solano, A. Marx, and R. Gross, Circuit quantum electrodynamics in the ultrastrong-coupling regime,
Nat. Phys. \textbf{6}, 772 (2010).

\bibitem{Forn10}
P. Forn-D\'{\i}az, J. Lisenfeld, D. Marcos, J. J. Garc\'{\i}a-Ripoll, E. Solano, C. J. P. M. Harmans, and J. E. Mooij,
Observation of the Bloch-Siegert Shift in a Qubit-Oscillator System in the Ultrastrong Coupling Regime,
Phys. Rev. Lett. \textbf{105}, 237001 (2010).

\bibitem{Forn16}
P. Forn-D\'{\i}az, J. J. Garc\'{\i}a-Ripoll, B. Peropadre, J.-L. Orgiazzi,
M. A. Yurtalan, R. Belyansky, C. M. Wilson, and A. Lupascu,
Ultrastrong coupling of a single artificial atom to an electromagnetic
continuum in the nonperturbative regime, Nat. Phys. {\bf 13}, 39 (2016).

\bibitem{You17}
Z. Chen, Y. Wang, T. Li, L. Tian, Y. Qiu, K. Inomata, F. Yoshihara, S. Han,
F. Nori, J. S. Tsai, and J. Q. You, Single-photon-driven high-order sideband transitions in an ultrastrongly coupled
circuit-quantum-electrodynamics system, Phys. Rev. A {\bf 96}, 012325 (2017).

\bibitem{Yoshihara16}
F. Yoshihara, T. Fuse, S. Ashhab, K. Kakuyanagi, S. Saito,
and K. Semba, Superconducting qubit-oscillator circuit beyond
the ultrastrong-coupling regime, Nat. Phys. {\bf 13}, 44 (2016).

\bibitem{Troiani18}
F. Troiani,
Readout of a weakly coupled qubit through the use of an auxiliary mode,
Phys. Lett. A {\bf 383}, 1536 (2019).


\bibitem{Bonizzoni18}
C. Bonizzoni, F. Troiani, A. Ghirri, and M. Affronte,
Microwave dual-mode resonators for coherent spin-photon coupling,
J. Appl. Phys. {\bf 124}, 194501 (2018).

\bibitem{Quijandria18}
F. Quijandr\'{\i}a, U. Naether, S. K. \"{O}zdemir, F. Nori, and D. Zueco,
$\mathcal{PT}$-symmetric circuit QED,
Phys. Rev. A \textbf{97}, 053846 (2018).

\bibitem{Bender98}
C. M. Bender and S. Boettcher,
Real Spectra in Non-Hermitian Hamiltonians Having $\mathcal{PT}$ Symmetry,
Phys. Rev. Lett. \textbf{80}, 5243 (1998).

\bibitem{Mostafazadeh02-1}
A. Mostafazadeh, Pseudo-Hermiticity versus $\mathcal{PT}$-symmetry: The necessary condition for the reality of the spectrum of a non-Hermitian Hamiltonian, J. Math. Phys. \textbf{43}, 205 (2002).

\bibitem{Mostafazadeh02-2}
A. Mostafazadeh, Pseudo-Hermiticity versus $\mathcal{PT}$-symmetry II: A complete characterization of non-Hermitian Hamiltonians with a real spectrum, J. Math. Phys. \textbf{43}, 2814 (2002).

\bibitem{Mostafazadeh02-3}
A. Mostafazadeh, Pseudo-Hermiticity versus $\mathcal{PT}$-symmetry III: Equivalence of pseudo-Hermiticity and the presence of antilinear symmetries, J. Math. Phys. \textbf{43}, 3944 (2002).

\bibitem{Bender05}
C. M. Bender,
Introduction to $\mathcal{PT}$-Symmetric Quantum Theory,
Contemp. Phys. \textbf{46}, 277 (2005).

\bibitem{Bender07}
C. M. Bender,
Making sense of non-Hermitian Hamiltonians,
Rep. Prog. Phys. \textbf{70}, 947 (2007).

\bibitem{Konotop16}
V. V. Konotop, J. Yang, and D. A. Zezyulin,
Nonlinear waves in $\mathcal{PT}$-symmetric systems,
Rev. Mod. Phys. \textbf{88}, 035002 (2016).

\bibitem{Heiss12}
W. D. Heiss,
The physics of exceptional points,
J. Phys. A \textbf{45}, 444016 (2012).

\bibitem{Lin11}
Z. Lin, H. Ramezani, T. Eichelkraut, T. Kottos, H. Cao, and D. N. Christodoulides,
Unidirectional Invisibility Induced by $\mathcal{PT}$-Symmetric Periodic Structures,
Phys. Rev. Lett. \textbf{106}, 213901 (2011).

\bibitem{Feng13}
L. Feng, Y. L. Xu, W. S. Fegadolli, M.-H. Lu, J. E. B. Oliveira, V. R. Almeida, Y. F. Chen, and A. Scherer,
Experimental demonstration of a unidirectional reflectionless parity-time metamaterial at optical frequencies,
Nat. Mater. \textbf{12}, 108 (2013).

\bibitem{Lv15}
X. Y. L\"{u}, H. Jing, J. Y. Ma, and Y. Wu,
$\mathcal{PT}$-symmetry-breaking chaos in optomechanics,
Phys. Rev. Lett. \textbf{114}, 253601 (2015).

\bibitem{Liertzer12}
M. Liertzer, L. Ge, A. Cerjan, A. D. Stone, H. E. T\"{u}reci, and S. Rotter,
Pump-Induced Exceptional Points in Lasers,
Phys. Rev. Lett. \textbf{108}, 173901 (2012).

\bibitem{Feng14}
L. Feng, Z. J. Wong, R. M. Ma, Y. Wang, and X. Zhang,
Single-mode laser by parity-time symmetry breaking,
Science \textbf{346}, 972 (2014).

\bibitem{Hodaei14}
H. Hodaei, M. A. Miri, M. Heinrich, D. N. Christodoulides, and M. Khajavikhan,
Parity-time-symmetric microring lasers,
Science \textbf{346}, 975 (2014).

\bibitem{Lin16}
Z. Lin, A. Pick, M. Lon\v{c}ar, and A. W. Rodriguez,
Enhanced Spontaneous Emission at Third-Order Dirac Exceptional Points in Inverse-Designed Photonic Crystals,
Phys. Rev. Lett. \textbf{117}, 107402 (2016).

\bibitem{Wiersig14}
J. Wiersig,
Enhancing the Sensitivity of Frequency and Energy Splitting Detection by Using Exceptional Points: Application to Microcavity Sensors for Single-Particle Detection,
Phys. Rev. Lett. \textbf{112}, 203901 (2014).

\bibitem{Liu16}
Z. P. Liu, J. Zhang, \c{S}. K. \"{O}zdemir, B. Peng, H. Jing, X. Y. L\"{u}, C. W. Li, L. Yang, F. Nori, and Y. X. Liu,
Metrology with $\mathcal{PT}$-Symmetric Cavities: Enhanced Sensitivity near the $\mathcal{PT}$-Phase Transition,
Phys. Rev. Lett. \textbf{117}, 110802 (2016).

\bibitem{Chen17}
W. Chen, \c{S}. K. \"{O}zdemir, G. Zhao, J. Wiersig, and L. Yang,
Exceptional points enhance sensing in an optical microcavity,
Nature (London) \textbf{548}, 192 (2017).

\bibitem{Hodaei17}
H. Hodaei, A. U. Hassan, S. Wittek, H. Garcia-Gracia, R. El-Ganainy, D. N. Christodoulides, and M. Khajavikhan,
Enhanced sensitivity at higher-order exceptional points,
Nature (London) \textbf{548}, 187 (2017).

\bibitem{Walls94}
D. F. Walls and G. J. Milburn, \emph{Quantum Optics} (Springer, Berlin, 1994).

\bibitem{Kurucz11}
Z. Kurucz, J. H. Wesenberg, and K. M{\o}lmer,
Spectroscopic properties of inhomogeneously broadened spin ensembles in a cavity,
Phys. Rev. A \textbf{83}, 053852 (2011).

\bibitem{Peropadre13}
B. Peropadre, D. Zueco, F. Wulschner, F. Deppe, A. Marx, R. Gross, and J. J. Garc\'{\i}a-Ripoll,
Tunable coupling engineering between superconducting resonators: From sidebands to effective gauge fields,
Phys. Rev. B \textbf{87}, 134504 (2013).

\bibitem{Baust15}
A. Baust, E. Hoffmann, M. Haeberlein, M. J. Schwarz, P. Eder, J. Goetz, F. Wulschner, E. Xie, L. Zhong, F. Quijandr\'{\i}a, B. Peropadre, D. Zueco, J.-J. Garc\'{\i}a Ripoll, E. Solano, K. Fedorov, E. P. Menzel, F. Deppe, A. Marx, and R. Gross,
Tunable and switchable coupling between two superconducting resonators,
Phys. Rev. B \textbf{91}, 014515 (2015).

\bibitem{Lee14}
T. E. Lee and C. K. Chan,
Heralded Magnetism in Non-Hermitian Atomic Systems,
Phys. Rev. X \textbf{4}, 041001 (2014).

\bibitem{Scheel18}
S. Scheel and A. Szameit,
$\mathcal{PT}$-symmetric photonic quantum systems with gain and loss do not exist,
Europhys. Lett. \textbf{122} 34001 (2018).

\bibitem{Kawabata17}
K. Kawabata, Y. Ashida, and M. Ueda,
Information Retrieval and Criticality in Parity-Time-Symmetric Systems,
Phys. Rev. Lett. \textbf{119}, 190401 (2017).


\end{thebibliography}
\end{document}